\documentclass[
prd
,showpacs ,amssymb,nobibnotes,aps,eqsecnum]{revtex4}
\usepackage{graphicx}
\input{epsf}

\usepackage{amsmath,amssymb}
\usepackage{bm}

\newcommand{\dalm}{\kern1pt\vbox{\hrule height 0.9pt\hbox{\vrule width 0.9pt
\hskip 2.5pt\vbox{\vskip 5.5pt}\hskip 3pt\vrule width 0.3pt}\hrule height 0.3pt}
\kern1pt}

\begin{document}



\title{Torsional oscillations in tensor-vector-scalar theory}

\author{Hajime Sotani} \email{hajime.sotani@nao.ac.jp}
\affiliation{
Division of Theoretical Astronomy, National Astronomical Observatory of Japan, 
2-21-1 Osawa, Mitaka, Tokyo 181-8588, Japan
}

\date{\today}

\begin{abstract}
With the Cowling approximation, the torsional oscillations on relativistic stars in tensor-vector-scalar (TeVeS) theory are examined. The  spectrum features in TeVeS are very similar to those in general relativity (GR), but the torsional frequencies in TeVeS become larger than those expected in GR. We find that, compared with the fluid oscillations with polar parity, the torsional frequencies depend strongly on the gravitational theory. Since the dependences of fundamental frequencies on the gravitational theory and on the equation of state are different from those of overtone, it could be possible to distinguish TeVeS from GR in the strong-field regime via observations of this type of oscillations with the help of the observation of stellar mass.
\end{abstract}

%
\maketitle
\section{Introduction}
\label{sec:I}

It is well-known that the theory of general relativity (GR) is valid in a weak gravitational field such as our Solar System, as has been shown via many experiments. Looking at the case in the strong gravitational field, the situation is quite different from the case in the weak-field regime, i.e., the tests of gravitational theory in the strong-field regime are still very poor. However, with the development of technology, it is becoming possible to observe compact objects with high accuracy, and with these observations it will be possible to test the gravitational theory in a strong gravitational field \cite{Psaltis2009}. In practice, one could observe compact objects not only via X-rays and $\gamma$-rays but also via gravitational waves emitted from the objects. So far, some possibilities for distinguishing the gravitational theory in the strong-field regime have been suggested. For instance, it has been proposed that one could distinguish the scalar-tensor theory \cite{Damour1992} from GR by using the surface atomic line redshifts \cite{DeDeo2003} or gravitational waves radiated from the neutron stars \cite{Sotani2004}. The possibility of a definitive test for GR with the direct observation of gravitational waves has also been pointed out \cite{Corda2009}.

As an alternative gravitational theory, tensor-vector-scalar (TeVeS) theory has attracted considerable attention in recent years, as proposed by Bekenstein \cite{Bekenstein2004} to extend the modified Newtonian dynamics \cite{Milgrom1983,Skordis2009} into a relativistic theory, i.e., a covariant theory. As well as the modified Newtonian dynamics, TeVeS can explain the galaxy rotational curve and Tully-Fisher law without the presence of dark matter \cite{Bekenstein2004}. This theory is also successful in explaining strong gravitational lensing \cite{Chen2006} and galaxy distributions through the evolving Universe without cold dark matter \cite{Dodelson2006}. For the strong gravitational region of TeVeS, Giannios found the Schwarzschild solution \cite{Giannios2005}, Sagi and Bekenstein found the Reissner-Nordstr\"{o}m solution \cite{Sagi2008}, and Lasky $et$ $al.$ produced the static, spherically symmetric stellar models in TeVeS by deriving the Tolman-Oppenheimer-Volkoff (TOV) equations in TeVeS \cite{Paul2008}. Additionally, Lasky and Donova examined the stability and quasinormal oscillations of the Schwarzshild solution in TeVeS \cite{Paul2010}. 

Furthermore, some have suggested how to distinguish TeVeS from GR observationally. For instance, one can reveal the gravitational theory in a strong gravitational field with the redshift of the atomic spectral line radiating from the neutron star surface \cite{Paul2008}, with Shapiro delays of gravitational waves and photons or neutrinos \cite{Desai2008}, with the spectrum of gravitational waves emitted from compact objects \cite{Sotani2009}, and with the rotational effect of neutron stars \cite{Sotani2010}. In fact, via observations of gravitational waves due to stellar oscillations, one can obtain information about the stellar parameters, such as mass, radius, rotational rate, magnetic fields, and equation of state (e.g., \cite{AK1996,Sotani2001,Sotani2003,Sotani2011a,Erich2009}), which is called ``gravitational wave asteroseismology." In addition to this, the direct detection of gravitational waves could possibly be used to determine the radius of the accretion disk around a supermassive black hole \cite{Sotani2006} or to know the magnetic effect during a stellar collapse \cite{Sotani2007a}. The test of gravitational theory in the strong-field regime is also one of the significant benefits of directly detecting gravitational waves.

In this article, we focus on the torsional oscillations in neutron stars, which are incompressible oscillations. Usually, the frequency of this type of oscillations should degenerate to zero if the neutron stars consist of matter without elasticity. However, realistic stellar models have the solid crust region near the stellar surface, and in this region shear torsional oscillations could exist. The boundary between the solid crust region and the fluid core is still uncertain because that depends strongly on the nuclear symmetric energy \cite{OI2007}, but the density of this boundary is proposed to be $2.4\times 10^{14}$ g/cm$^3$ \cite{NV} or $1.28\times 10^{14}$ g/cm$^3$ \cite{DH}. In this article, we adopt the density of $2.0\times 10^{14}$ g/cm$^3$ to see the dependence of torsional frequencies on the gravitational theory. From the observational point of view, the observed quasiperiodic oscillations in giant flares \cite{WS2006} are considered to come from this type of oscillations \cite{Sotani2007}.

In order to see the dependence of torsional frequencies on the gravitational theory, we adopt the relativistic Cowling approximation as a first step in this article, i.e., the perturbations except for the fluid one are neglected. It is known that this approximation is quite good at least in GR, because the torsional oscillations do not involve the density variation. A more detailed study including the perturbations of other fields will be done in the near future.

This article is organized as follows. In the next section, we mention the stellar model in TeVeS, and in Sec. \ref{sec:III} we derive the perturbation equations for torsional oscillations with the Cowling approximation. In Sec. \ref{sec:IV}, the concrete calculations of the obtained perturbation equations will be performed to show the dependence on the gravitational theory. Finally, we conclude in Sec. \ref{sec:V}. In this article, we adopt the unit of $c=G=1$, where $c$ and $G$ denote the speed of light and the gravitational constant, respectively, and the metric signature is $(-,+,+,+)$.

\section{Stellar Models in TeVeS}
\label{sec:II}

In this section, we only mention the fundamental parts of the theory. (See \cite{Bekenstein2004} for details of TeVeS.) TeVeS is based on three dynamical gravitational fields: an Einstein metric $g_{\mu\nu}$, a timelike four-vector field ${\cal U}^\mu$, and a scalar field $\varphi$, in addition to a nondynamical scalar field $\sigma$. The vector field fulfills the normalization condition with the Einstein metric as $g_{\mu\nu}{\cal U}^\mu{\cal U}^\nu=-1$, and the physical metric $\tilde{g}_{\mu\nu}$ is defined as
\begin{equation}
 \tilde{g}_{\mu\nu} = e^{-2\varphi}g_{\mu\nu} - 2{\cal U}_\mu{\cal U}_\nu\sinh(2\varphi).
\end{equation}
All quantities in the physical frame are denoted with a tilde, and any quantity without a tilde is in the Einstein frame. The total action of TeVeS, $S$, contains contributions from the three dynamical fields mentioned above as well as the matter contribution \cite{Bekenstein2004}, which includes two positive dimensionless parameters, $k$ and $K$, corresponding to the coupling parameters for the scalar and vector fields. The field equations for the tensor, vector, and scalar fields can be obtained by varying the total action with respect to $g_{\mu\nu}$, ${\cal U}^\mu$, and $\varphi$, respectively. (See \cite{Bekenstein2004} for the explicit forms.) Since a previous study for the neutron star structures in TeVeS has shown that the stellar properties are almost independent from the scalar coupling $k$ \cite{Paul2008}, in this article we focus only on the dependence of vector coupling $K$. The restriction on $K$ has not been discussed in great detail in the literature, but the authors of\cite{Paul2008} mentioned that $K$ has to be less than 2 to construct the stellar models, and also that $K$ should be less than 1 to produce a realistic stellar mass. According to this restriction on $K$, in this article we examine $K$ as varying in the range of $0\le K\le 1$. It should be noticed that GR is corresponding to the special case that $K=0$ and $k=0$ in TeVeS.

As a background stellar model, we consider nonrotating relativistic stars, which have been investigated in \cite{Paul2008}. A static, spherically symmetric stellar model can be expressed with the following metric:
\begin{equation}
 ds = -e^\nu dt^2 + e^\zeta dr^2 + r^2 \left(d\theta^2 + \sin^2\theta d\phi^2\right),
\end{equation}
where $\nu$ and $\zeta$ are functions of the radial coordinate $r$. Although the vector field on a static, spherically symmetric spacetime can be generally described as ${\cal U}^\mu=\left({\cal U}^t,{\cal U}^r,0,0\right)$, for simplicity we adopt in this article the simple case that the radial component should be zero, i.e., ${\cal U}^r=0$, which is the same assumption adopted in \cite{Paul2008,Sotani2009,Sotani2010}. Then, with the normalization condition, the vector field can be fully determined as ${\cal U}^\mu=(e^{-\nu/2},0,0,0)$. With this vector field, the physical metric is
\begin{equation}
 d\tilde{s}^2 = -e^{\nu+2\varphi}dt^2 + e^{\zeta-2\varphi}dr^2 + r^2 e^{-2\varphi}\left(d\theta^2 + \sin^2\theta d\phi^2\right),
\end{equation}
and the fluid four-velocity is $\tilde{u}_\mu=e^{\varphi}{\cal U}_\mu$. Regarding the stellar matter, we assume a perfect fluid described by the energy-momentum tensor
\begin{equation}
  \tilde{T}_{\mu\nu}= \left(\tilde{p}+\tilde{\rho}\right)\tilde{u}_\mu\tilde{u}_\nu+\tilde{p}\tilde{g}_{\mu\nu},
\end{equation}
where $\tilde{p}$ and $\tilde{\rho}$ are the pressure and energy density in the physical frame, respectively. Furthermore, in order to construct the stellar model, one needs to prepare the relation between $\tilde{p}$ and $\tilde{\rho}$, i.e., the equation of state (EOS). In this article, we adopt the same EOS's as in \cite{Sotani2004}, which are polytropic ones derived by fitting functions to tabulated data of realistic EOS's known as EOS A and EOS II.  With these EOS's, the maximum masses of a neutron star in GR are $M=1.65M_\odot$ for EOS A and $M=1.95M_\odot$ for EOS II. That is, EOS A and EOS II are considered as soft and intermediate EOS's, respectively. At last, the stellar modes in TeVeS can be constructed by using the recipe shown in \cite{Paul2008}.

\section{Torsional Oscillations}
\label{sec:III}

As mentioned above, in this article, we focus on the torsional oscillations with the relativistic Cowling approximation, where we consider only the fluid perturbation with axial parity and the other perturbations of vector and tensor fields are neglected. It should be noted that the perturbation of scalar field exists only for polar perturbation. 
Considering the torsional oscillations, the Lagrangian displacement vector for the fluid perturbation can be expressed as
\begin{equation}
 \tilde{\xi}^{i} = \left(0,0,Z(t,r)\frac{1}{\sin\theta}\partial_\theta P_\ell\right),
\end{equation}
where $\partial_\theta$ denotes the partial derivative with respect to $\theta$, while $P_\ell=P_\ell(\cos\theta)$ is the Legendre polynomial of order $\ell$. Then, the nonzero component of the perturbed fluid four-velocity in the physical frame can be written as
\begin{equation}
 \delta\tilde{u}^\phi = e^{-\varphi-\nu/2}\partial_t Z \frac{1}{\sin\theta}\partial_\theta P_\ell,
\end{equation}
where $\partial_t$ denotes the partial derivative with respect to $t$. The perturbed energy-momentum tensor including the contribution from shear is given by
\begin{equation}
 \delta\tilde{T}_{\mu\nu} = \left(\tilde{p}+\tilde{\rho}\right)\left(\delta\tilde{u}_\mu \tilde{u}_\nu + \tilde{u}_\mu\delta \tilde{u}_\nu\right)
     - 2\tilde{\mu}\delta \tilde{S}_{\mu\nu},
\end{equation}
where $\tilde{\mu}$ is the shear modulus and $\tilde{S}_{\mu\nu}$ is the shear tensor defined via $\tilde{\sigma}_{\mu\nu}={\cal L}_{\tilde{u}}\tilde{S}_{\mu\nu}$ \cite{CQ1973,ST1983}. Here $\tilde{\sigma}_{\mu\nu}$ is the rate of shear tensor, which is defined as
\begin{equation}
 \tilde{\sigma}_{\mu\nu} = \frac{1}{2}\left(\tilde{P}^{\alpha}_{\ \nu}\tilde{\nabla}_\alpha\tilde{u}_{\mu}
     + \tilde{P}^{\alpha}_{\ \mu}\tilde{\nabla}_\alpha\tilde{u}_{\nu}\right)
     - \frac{1}{3}\tilde{P}_{\mu\nu}\tilde{\nabla}_\alpha\tilde{u}^{\alpha},
\end{equation}
where $\tilde{P}_{\mu\nu}$ is the projection tensor
\begin{equation}
 \tilde{P}_{\mu\nu} = \tilde{g}_{\mu\nu} + \tilde{u}_\mu\tilde{u}_\nu.
\end{equation}
The speed of shear waves can be expressed as $\tilde{v}^2_s=\tilde{\mu}/(\tilde{p}+\tilde{\rho})$ \cite{ST1983}, where a typical value of $\tilde{v}_s$ is around $10^8$ cm/s in the crust of neutron stars. In this article, we adopt the shear modulus inside the crust determined by this simple relation. Finally, one can get the equations describing the fluid perturbations by taking a variation of the energy-momentum conservation law, i.e., $\delta(\tilde{\nabla}_\beta\tilde{T}^{\alpha\beta})=0$, which reduces to $\tilde{\nabla}_\beta\delta \tilde{T}^{\alpha\beta}=0$ with the Cowling approximation. The explicit form with $\alpha=\phi$ becomes
\begin{equation}
 \left(\tilde{p}+\tilde{\rho}\right)e^{-\nu-4\varphi}\ddot{Z} - \tilde{\mu}e^{-\zeta}Z'' 
      - \left[\tilde{\mu}'+\left(\frac{4}{r}-2\varphi'+\frac{\nu'-\zeta'}{2}\right)\tilde{\mu}\right]e^{-\zeta}Z'
      + \frac{(\ell+2)(\ell-1)}{r^2}\tilde{\mu}Z=0, \label{eq:perturbation}
\end{equation}
where $\dot{(\ )}$ and $(')$ denote the partial derivative with respect to $t$ and $r$, respectively. Assuming that the perturbed variable has a harmonic time dependence, such as $Z(t,r)=Z(r)e^{i\omega t}$, Eq. (\ref{eq:perturbation}) reduces to
\begin{equation}
 \tilde{\mu}Z'' + \left[\tilde{\mu}'+\left(\frac{4}{r}-2\varphi'+\frac{\nu'-\zeta'}{2}\right)\tilde{\mu}\right]Z'
      + \left[\left(\tilde{p} + \tilde{\rho}\right)\omega^2e^{-\nu-4\varphi} - \frac{(\ell+2)(\ell-1)}{r^2}\tilde{\mu}\right]e^\zeta Z = 0.
\end{equation}
Imposing appropriate boundary conditions on this equation, the problem to solve becomes the eigenvalue problem. In practice, the above equation will be integrated only in the crust region, i.e., the boundary conditions are imposed at the basis of crust ($r=R_c$) and at the stellar surface $(r=R)$, because $\tilde{\mu}=0$ in the fluid core as mentioned in the Introduction. In this article, we impose a zero traction condition at $r=R_c$ and the zero-torque condition at $r=R$ \cite{ST1983}. Both conditions correspond to $Z'=0$.

\section{Oscillation Spectra}
\label{sec:IV}

\begin{figure}[htbp]
\begin{center}
\begin{tabular}{cc}
\includegraphics[scale=0.45]{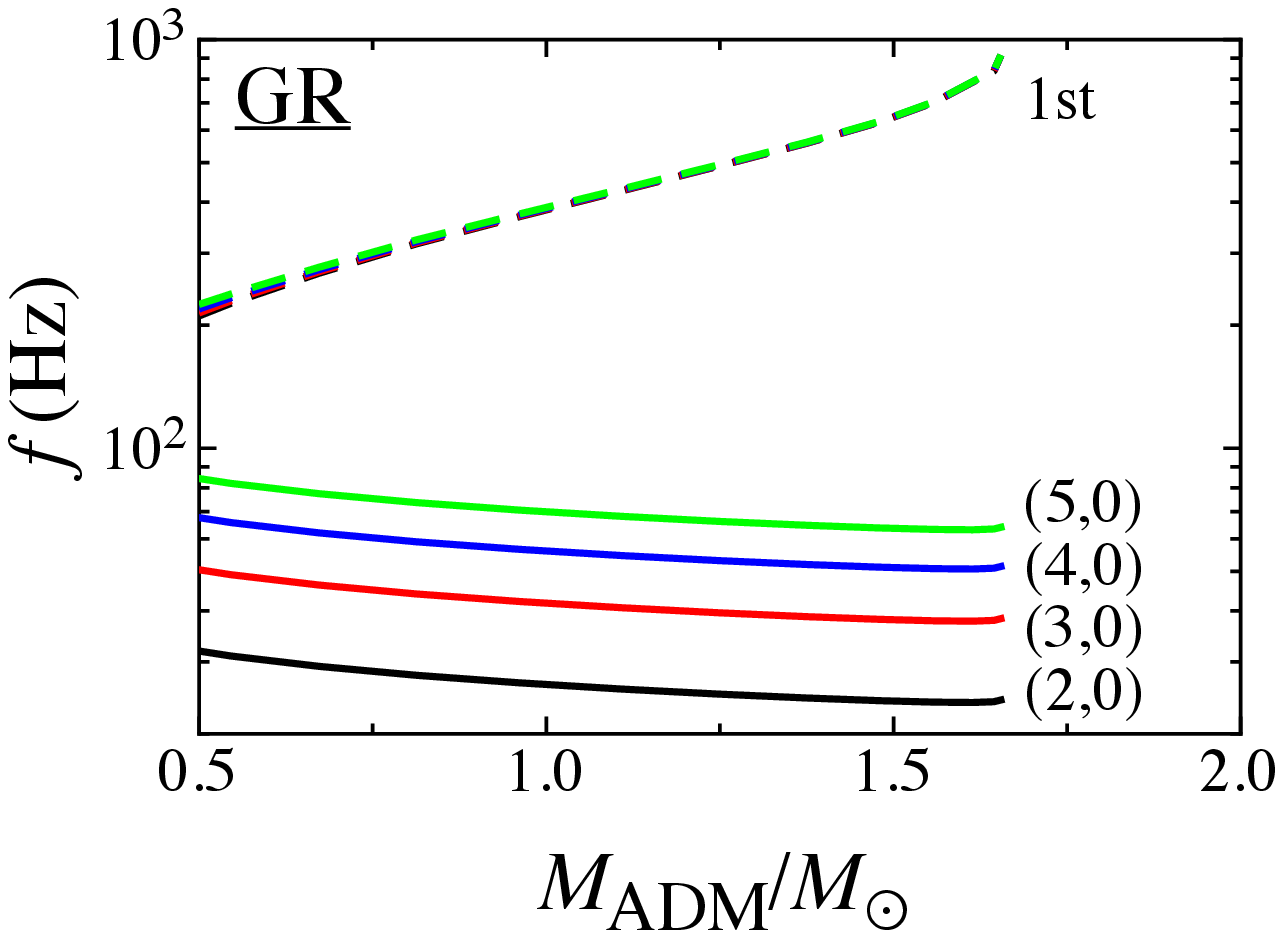} &
\includegraphics[scale=0.45]{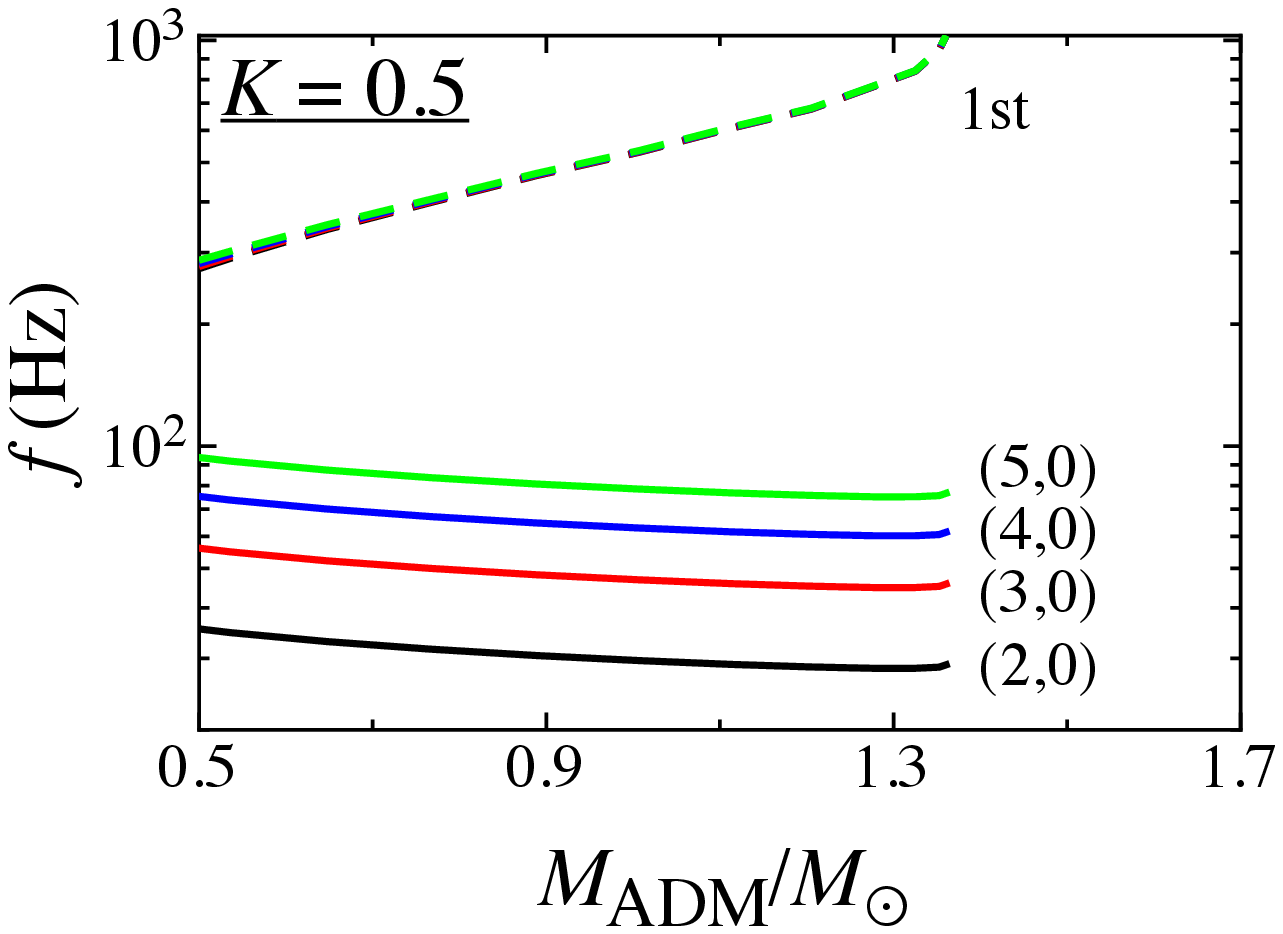} \\
\end{tabular}
\end{center}
\caption{
Frequencies of fundamental modes and first overtone are plotted as functions of the ADM mass with EOS A in GR (left panel) and in TeVeS with $K=0.5$ (right panel). In the figures, the solid lines correspond to the fundamental frequencies with $\ell=2$, 3, 4, and $5$, i.e., $(\ell,n)=(2,0)$, (3,0), (4,0), and (5,0), where $n$ is the number of nodes in the eigenfunctions, while the broken lines correspond to the first overtone with $\ell=2$, 3, 4, and $5$.
}
\label{fig:t2}
\end{figure}

In this section, we examine the torsional oscillations in the crust region of neutron stars both in GR and in TeVeS. With respect to the torsional modes in GR, it is known that one can see the dependence of $\ell$ in the fundamental modes, while the frequencies of overtone are almost independent of $\ell$, as shown in the left panel of Fig. \ref{fig:t2}, where the specific frequencies in GR are plotted as a function of the stellar mass. On the other hand, we have done the numerical calculations to determine the frequencies in TeVeS with different value of $K$. As an example, the result with $K=0.5$ is shown in the right panel of Fig. \ref{fig:t2}. From this figure, it is shown that the feature of torsional modes in GR can be kept even in TeVeS.

\begin{figure}[htbp]
\begin{center}
\begin{tabular}{cc}
\includegraphics[scale=0.45]{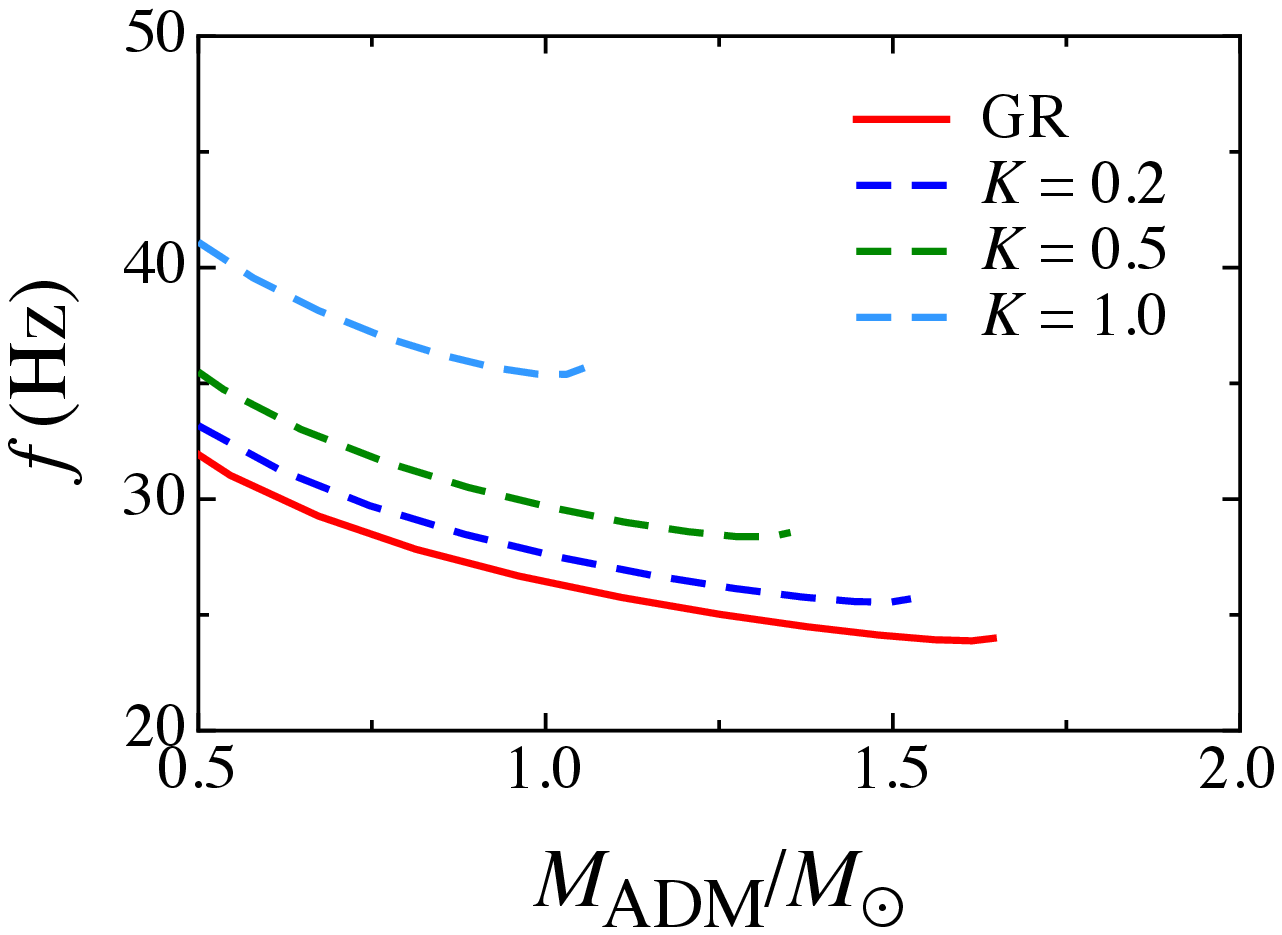} &
\includegraphics[scale=0.45]{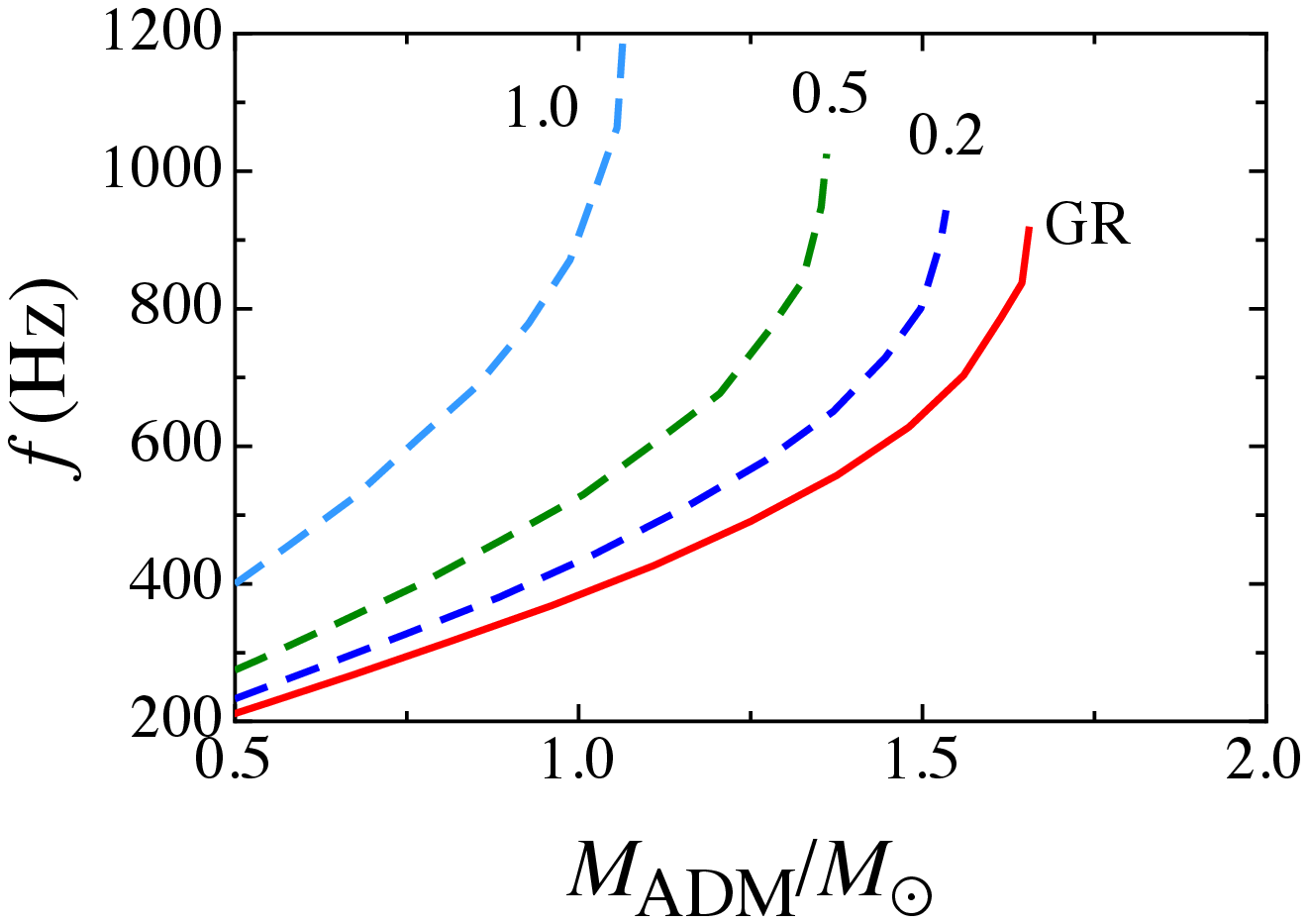} \\
\end{tabular}
\end{center}
\caption{
Frequencies of fundamental modes and first overtone with $\ell=2$ are plotted as functions of the ADM mass with EOS A. Left and right panels correspond to the fundamental and first overtone frequencies, respectively. In both figures, the solid and broken lines correspond to the frequencies in GR and TeVeS, respectively.
}
\label{fig:t2-A}
\end{figure}

\begin{figure}[htbp]
\begin{center}
\begin{tabular}{cc}
\includegraphics[scale=0.45]{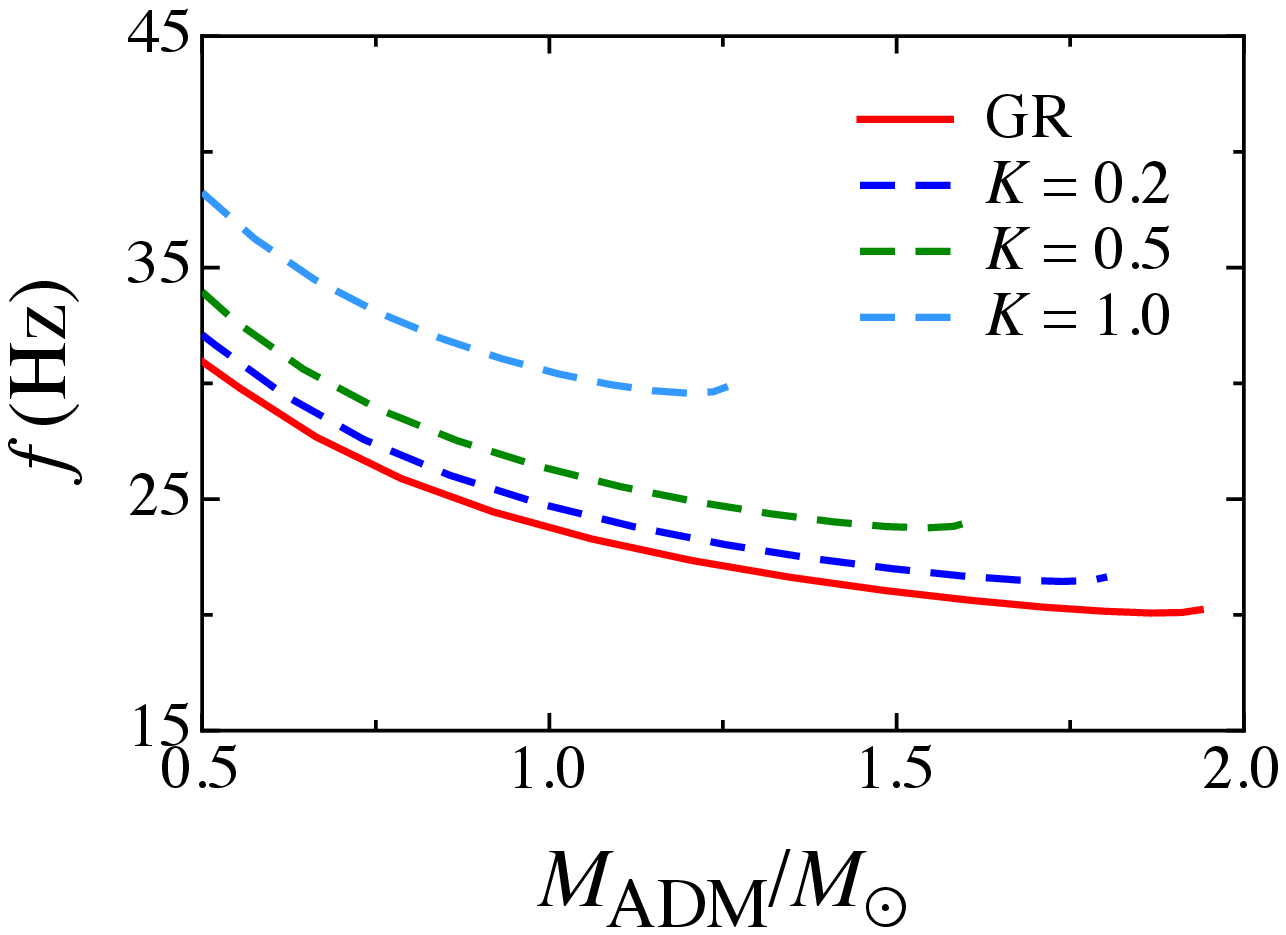} &
\includegraphics[scale=0.45]{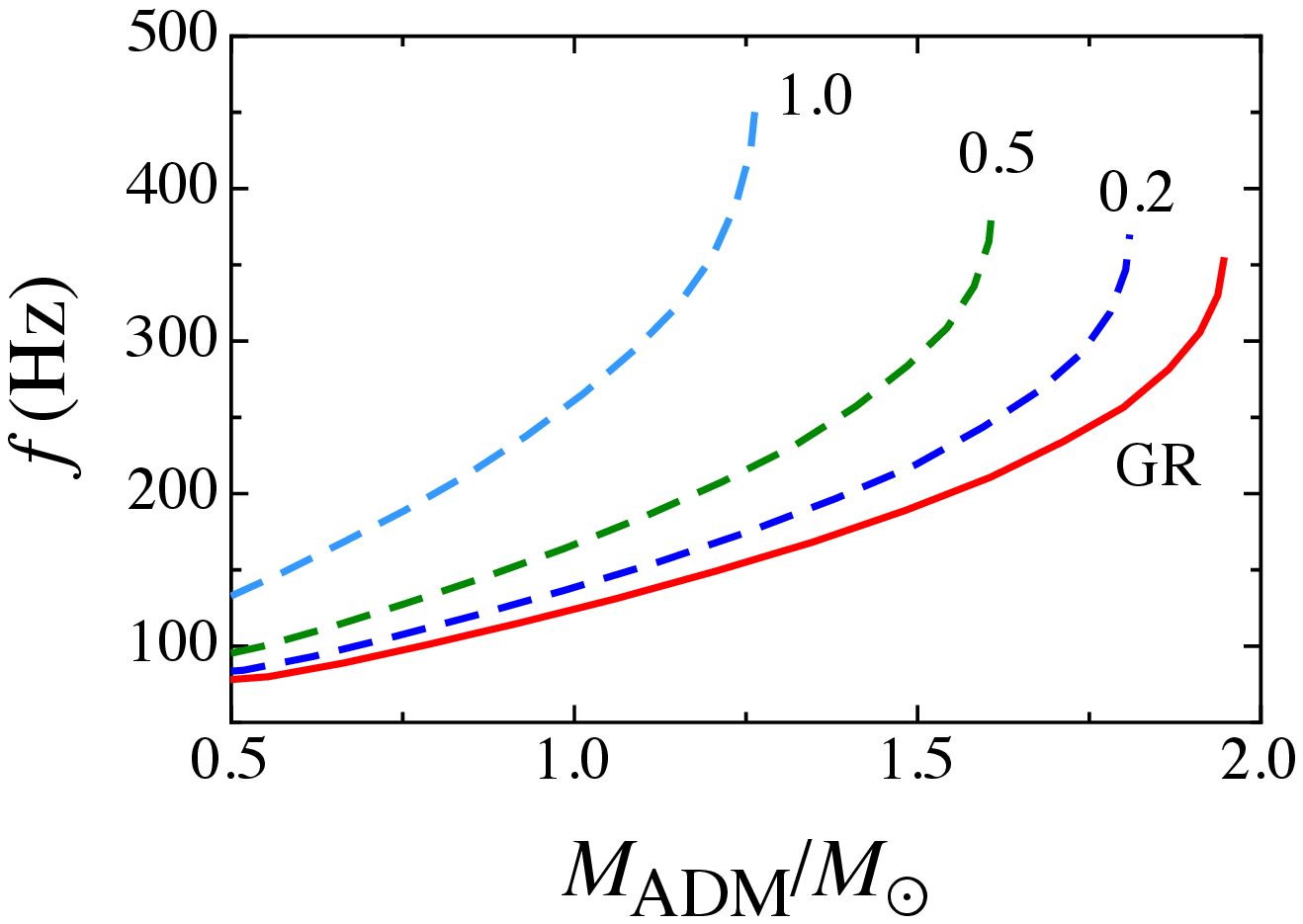} \\
\end{tabular}
\end{center}
\caption{
Similar to Fig. \ref{fig:t2-A}, but with EOS II.
}
\label{fig:t2-II}
\end{figure}

In order to see the dependence of frequencies of torsional oscillations on the gravitational theory, we focus on the $\ell=2$ oscillation modes. Figure \ref{fig:t2-A} shows the fundamental frequencies (left panel) and the frequencies of first overtone (right panel), in both GR (solid lines) and TeVeS (broken lines) for the stellar models with EOS A. Similar to Fig. \ref{fig:t2}, the frequencies are shown as functions of the stellar mass. From this figure, one can easily observe that the frequencies expected in TeVeS are quite different from those in GR. In practice, depending on the value of $K$, the frequencies in TeVeS become $34\%$ larger for the fundamental oscillations and $150\%$ larger for the first overtone than those in GR. In Fig. \ref{fig:t2-II}, we draw a figure similar to Fig. \ref{fig:t2-A} but EOS II. One can observe that the relative augmentation of the frequencies in TeVeS compared with those in GR becomes $28\%$ for the fundamental oscillations and $116\%$ for the first overtone. That is, the stellar models with softer EOS seem to be more sensitive about the gravitational theory than those with stiffer EOS. Additionally, it should be emphasized that, compared with the fluid oscillations of polar parity accompanied by variation of density, the frequencies of torsional modes depend strongly on the gravitational theory, where the frequencies of polar parity in TeVeS become around $20\%$ larger than those expected in GR \cite{Sotani2009}. And, as mentioned in \cite{Sotani2009}, since this deviation between the frequencies in GR and in TeVeS results from the existence of a scalar field, observing the torsional oscillations as well as the fluid oscillations of polar parity could tell us the existence of the scalar field.

%
%

\begin{figure}[htbp]
\begin{center}
\begin{tabular}{cc}
\includegraphics[scale=0.45]{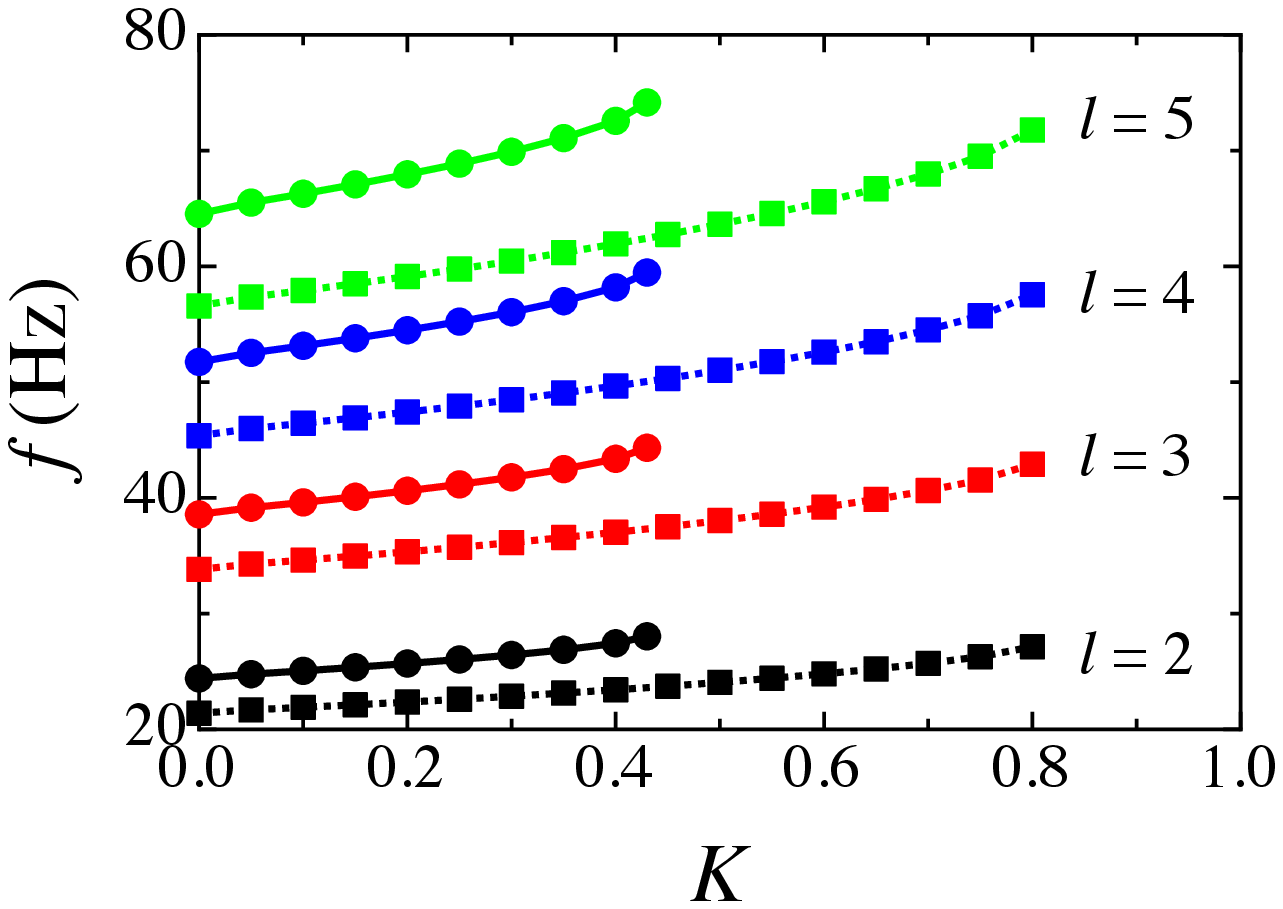} &
\includegraphics[scale=0.45]{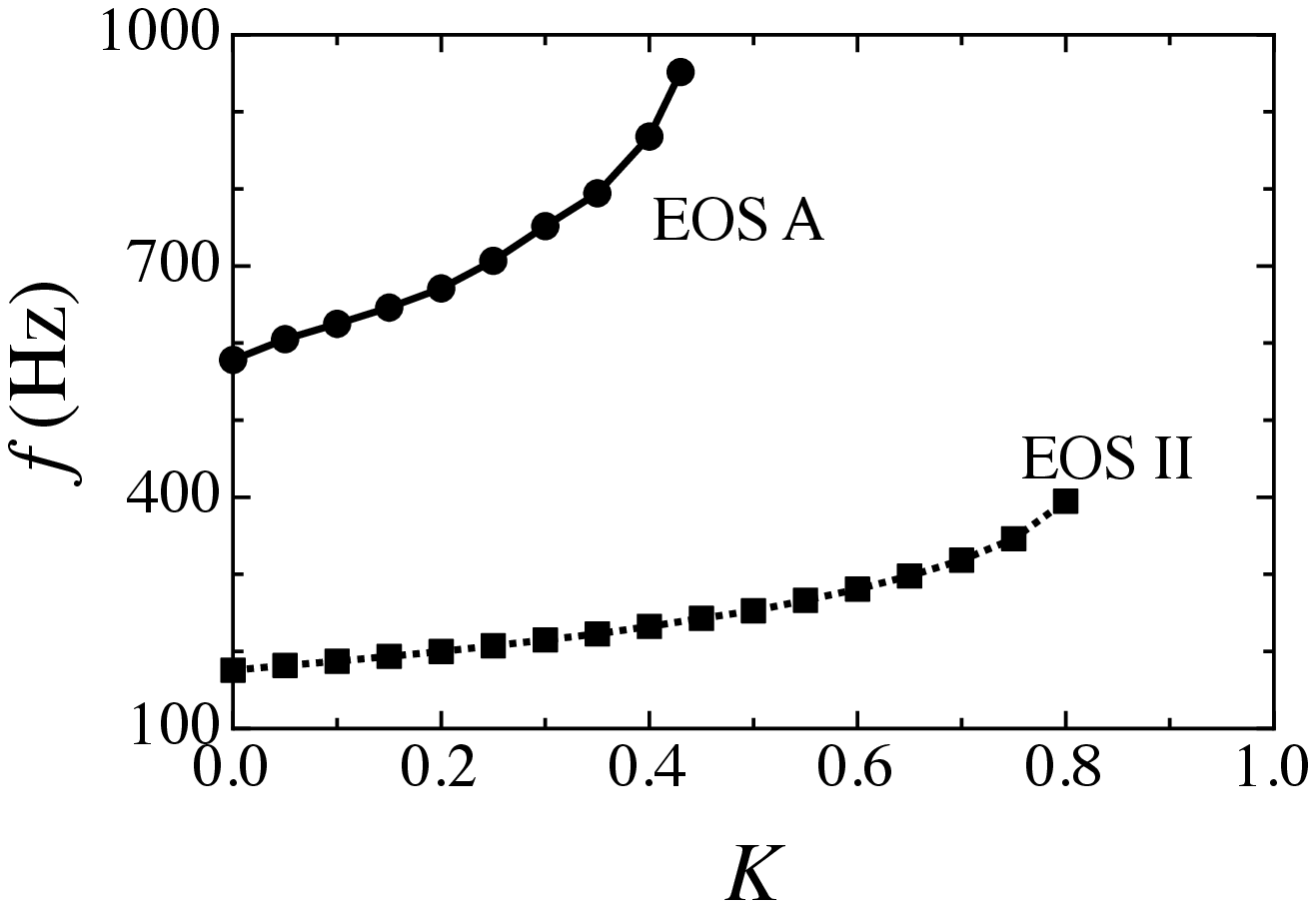} \\
\end{tabular}
\end{center}
\caption{
For stellar models with $M_{\rm ADM}=1.4M_\odot$, the frequencies of fundamental and first overtone torsional modes are shown as functions of parameter $K$ with EOS A (solid lines with circles) and EOS II (dotted lines with squares). The left panel corresponds to the fundamental modes with $\ell=2$, 3, 4, and 5, while the right panel corresponds to the first overtone with $\ell=2$.
}
\label{fig:t-K}
\end{figure}

Furthermore, in Fig. \ref{fig:t-K}, we plot the fundamental frequencies (left panel) and the first overtone (right panel) of torsional modes as functions of parameter $K$, where the ADM masses are fixed to be $1.4M_\odot$. Here, the frequencies with $\ell=2$, 3, 4, and 5 are shown in the left panel, while those with only $\ell=2$ are shown in the right panel, because the frequencies of overtone are almost independent of $\ell$, as we saw in Fig. \ref{fig:t2}. Additionally, in both panels, the results for EOS A and EOS II are plotted using the solid lines with circle marks and the dotted lines with square marks, respectively, and the results in GR are shown at $K=0$. From these figures, one can observe  that the qualitative dependencies of frequency on the value of $K$ are independent of the adopted EOS and the value of $\ell$. That is, the frequencies of torsional modes are increasing as the value of $K$ becomes large. From the left panel of Fig. \ref{fig:t-K}, since the dependence of $K$ on the fundamental frequencies is almost comparable to that of EOS, it might be difficult to distinguish the gravitational theory by using only the observation of fundamental modes. On the other hand, the right panel of Fig. \ref{fig:t-K} shows that the frequencies of overtone depend strongly on EOS rather than on the value of $K$. So, with observation of the frequencies of overtone, it could be possible to make a constraint in EOS independent of the gravitational theory. Then, after making a constraint in EOS with the help of observation of the stellar mass, it might be possible to restrict on the value of $K$ using the observations of fundamental oscillations, i.e., it might be possible to probe the gravitational theory in the strong-field regime observationally.

%
%

\section{Conclusion}
\label{sec:V}

In this article, focusing on the torsional oscillations inside the crust region, we have derived the perturbation equation of neutron stars in TeVeS and calculated their eigenfrequencies. We find that the frequencies of torsional modes in TeVeS become larger than those expected in GR, whose dependence on the gravitational theory is stronger than that of fluid oscillations involving the variation of density. It is also found that the stellar models with softer EOS are more sensitive about the gravitational theory than those with stiffer EOS. Since the frequencies of overtone depend strongly on the adopted EOS rather than on the gravitational theory, it is possible, via the observation of the torsional frequencies of overtone, to restrict on the EOS independent of the gravitational theory. Then, with the help of the observation of stellar mass, it might also be possible to make a constraint in the gravitational theory via the observation of fundamental torsional modes. It should be noticed that, since these imprints of TeVeS come from the presence of a scalar field, it could be also possible, via observations of such oscillations of neutron star, to probe the existence of a scalar field.

As a first step, we assume the Cowling approximation in this article, i.e., our examinations are restricted to only fluid oscillations, so we should make a more detailed study including the metric and vector field perturbations. Via these oscillations, one could obtain additional information, and combining that with the results reported in this article would provide more accurate constraints on the gravitational theory in the strong-field regime. Additionally, since it is suggested that the nuclear structure in the bottom of a crust could be nonuniform, the so-called ``pasta structure," we might consider the effect of the pasta phase on the torsional oscillations \cite{Sotani2011b}. Furthermore, we should take into account the magnetic effects on the frequencies, because the torsional oscillation could be detected in the magnetars as the quasiperiodic oscillations during the giant flares \cite{Sotani2008}. Considering these additional effects, one must obtain more accurate constraints in a more realistic situation.

\acknowledgments
This work was supported by Grant-in-Aid for Scientific Research on Innovative Areas (23105711).




\begin{thebibliography}{999}

\bibitem{Psaltis2009}
    D. Psaltis,
    Living Rev. Relativity {\bf 11}, 001 (2009).

\bibitem{Damour1992}
    T. Damour and G. Esposito-Far{\` e}se,
    Classical Quantum Gravity {\bf 9}, 2093 (1992).

\bibitem{DeDeo2003}
    S. DeDeo and D. Psaltis,
    Phys. Rev. Lett. {\bf 90}, 141101 (2003).

\bibitem{Sotani2004}
    H. Sotani and K.D. Kokkotas,
    Phys. Rev. D {\bf 70}, 084026 (2004);
    {\bf 71}, 124038 (2005).

\bibitem{Corda2009}
    C. Corda,
    Int. J. Mod. Phys. D {\bf 18}, 2275 (2009).

\bibitem{Bekenstein2004}
    J.D. Bekenstein,
    Phys. Rev. D {\bf 70}, 083509 (2004).

\bibitem{Milgrom1983}
	M. Milgrom,
	Astrophys. J. {\bf 270}, 365 (1983).

\bibitem{Skordis2009}
   C. Skordis,
   Class. Q. Grav. {\bf 26}, 143001 (2009).

\bibitem{Chen2006}
    D.M. Chen and H.S. Zhao,
    Astrophys. J. {\bf 650}, L9 (2006).

\bibitem{Dodelson2006}
	S. Dodelson and M. Liguori,
	Phys. Rev. Lett. {\bf 97}, 231301 (2006).

\bibitem{Giannios2005}
    D. Giannios,
    Phys. Rev. D {\bf 71}, 103511 (2005).

\bibitem{Sagi2008}
    E. Sagi and J.D. Bekenstein,
    Phys. Rev. D {\bf 77}, 024010 (2008).

\bibitem{Paul2008}
    P.D. Lasky, H. Sotani, and D. Giannios,
    Phys. Rev. D {\bf 78}, 104019 (2008).

\bibitem{Paul2010}
    P.D. Lasky and D.D. Doneva,
    Phys. Rev. D {\bf 82}, 124068 (2010).

\bibitem{Desai2008}
    S. Desai, E.O. Kahya, and R.P. Woodard,
    Phys. Rev. D {\bf 77}, 124041 (2008).

\bibitem{Sotani2009}
    H. Sotani,
    Phys. Rev. D {\bf 79}, 064033 (2009); {\bf 80} 064035 (2009).

\bibitem{Sotani2010}
    H. Sotani,
    Phys. Rev. D {\bf 81}, 084006 (2010); {\bf 82}, 124061 (2010).

\bibitem{AK1996}
   N. Andersson and K.D. Kokkotas,
   Phys. Rev. Lett. {\bf 77}, 4134 (1996).

\bibitem{Sotani2001}
   H. Sotani, K. Tominaga, and K.I. Maeda,
   Phys. Rev. D {\bf 65}, 024010 (2001).

\bibitem{Sotani2003}
   H. Sotani and T. Harada,
   Phys. Rev. D {\bf 68}, 024019 (2003);
   H. Sotani, K. Kohri, and T. Harada,
   $ibid$. {\bf 69}, 084008 (2004).

\bibitem{Sotani2011a}
   H. Sotani, N. Yasutake, T. Maruyama, and T. Tatsumi,
   Phys. Rev. D {\bf 83}, 024014 (2011).


\bibitem{Erich2009}
    E. Gaertig and K.D. Kokkotas,
    Phys. Rev. D {\bf 80}, 064026 (2009).

\bibitem{Sotani2006}
   H. Sotani and M. Saijo,
   Phys. Rev. D {\bf 74}, 024001 (2006).

\bibitem{Sotani2007a}
   H. Sotani, S. Yoshida, and K.D. Kokkotas,
   Phys. Rev. D {\bf 75}, 084015 (2007);
   H. Sotani,
   $ibid.$ {\bf 79}, 084037 (2009).

\bibitem{OI2007}
    K. Oyamatsu and K. Iida,
    Phys. Rev. C {\bf 75}, 015801 (2007).

\bibitem{NV}
    J.W. Negele and D. Vautherin,
    Nucl. Phys. A {\bf 207}, 298 (1973).

\bibitem{DH}
    F. Douchin and P. Haensel,
    A\&A {\bf 380}, 151 (2001).





\bibitem{WS2006}
    A.L. Watts and T.E. Strohmayer,
    Adv. Space Res. {\bf 40}, 1446 (2006).

\bibitem{Sotani2007}
    H. Sotani, K.D. Kokkotas, and N.Stergioulas,
    Mon. Not. R. Astron Soc. {\bf 375}, 261 (2007).

\bibitem{CQ1973}
    B. Carter and H. Quintana,
    Proc. R. Soc. Lond. A {\bf 331}, 57 (1973).

\bibitem{ST1983}
    B.L.Schumaker and K.S.Thorne,
    Mon. Not. R. Astron Soc. {\bf 203}, 457 (1983)

\bibitem{Sotani2011b}
    H. Sotani, arXiv:1106.2621.

\bibitem{Sotani2008}
    H. Sotani, K.D. Kokkotas, and N.Stergioulas,
    Mon. Not. R. Astron Soc. {\bf 385}, L5 (2008);
    H. Sotani, A. Colaiuda, and K.D. Kokkotas,
    $ibid$. {\bf 385}, 2161 (2008);
    H. Sotani and K.D. Kokkotas,
    $ibid$. {\bf 395}, 1163 (2009).



\end{thebibliography}
\end{document}